\documentclass[%
reprint,
superscriptaddress,
aps,
prb,
]{revtex4-2}

\usepackage{graphicx}
\usepackage{dcolumn}
\usepackage{hyperref}
\usepackage{url}
\usepackage{bm}
\usepackage{pifont}
\usepackage{xcolor}

\begin{document}


\title{Structural Phase Separation and Enhanced Superconductivity in  $\mathbf{La}_{1.875}\mathbf{Ba}_{0.125}\mathbf{CuO}_{4}$ under Uniaxial Strain}

\author{Baizhi Gao}
 \affiliation{Department of Physics, University of Toronto, Toronto, Ontario M5S 1A7, Canada}
\author{Ehsan Nikbin}
 \affiliation{Department of Materials Science and Engineering, University of Toronto, Toronto, Ontario M5S 3E4, Canada}
\author{Graham Johnstone}
 \affiliation{Department of Physics, University of Toronto, Toronto, Ontario M5S 1A7, Canada}
\author{Ze Shi}
 \affiliation{Department of Physics, University of Toronto, Toronto, Ontario M5S 1A7, Canada}
\author{Christopher Heath}
 \affiliation{Department of Physics, University of Toronto, Toronto, Ontario M5S 1A7, Canada}
\author{Narayan Appathurai}
 \affiliation{Canadian Light Source, University of Saskatchewan, Saskatoon, Saskatchewan S7N 0X4, Canada}
\author{Beatriz Diaz Moreno}
 \affiliation{Canadian Light Source, University of Saskatchewan, Saskatoon, Saskatchewan S7N 0X4, Canada}
\author{Al Rahemtulla}
 \affiliation{Canadian Light Source, University of Saskatchewan, Saskatoon, Saskatchewan S7N 0X4, Canada}
\author{G. D. Gu}
 \affiliation{Condensed Matter Physics and Materials Science Department, Brookhaven National Laboratory, Upton, New York 11973, USA}
\author{John M. Tranquada}
 \affiliation{Condensed Matter Physics and Materials Science Department, Brookhaven National Laboratory, Upton, New York 11973, USA}
\author{Jane Y. Howe}
 \affiliation{Department of Materials Science and Engineering, University of Toronto, Toronto, Ontario M5S 3E4, Canada}
\author{Young-June Kim}
\email[Email: ]{youngjune.kim@utoronto.ca} 
\affiliation{Department of Physics, University of Toronto, Toronto, Ontario M5S 1A7, Canada}


\begin{abstract}
Strain engineering has attracted significant attention in recent years due to its capability in tuning lattice and electronic structures of quantum materials. Using moderate uniaxial compressive strain, we induce structural phase separation in the low temperature phase of x=1/8 $\rm La_{2-x}Ba_{x}CuO_{4}$ (LBCO) single crystal. These structures are low temperature tetragonal (LTT), low temperature less orthorhombic (LTLO), and a plastically deformed nano-domain structure (PDNS), comprised of few-nanometer-sized orthorhombic domains within amorphous matrix. These three structures exhibit distinct superconducting behaviors. The volume fraction of the LTT structure is suppressed with increasing strain, while its superconducting transition temperature increases and broadens. The LTLO structure exhibits a sharp superconducting transition above $32$~K, which increases up to $\sim 36$~K at maximum strain. The PDNS phase exhibits a very broad superconducting transition, persisting even after removing the strain. Our study illustrates the sensitivity of superconductivity to the structure of the LBCO sample near its stripe instability.

\end{abstract}

\maketitle


\section{\label{sec:level1}Introduction}

With the goal of developing novel materials and uncovering new physics, strain engineering of quantum materials has attracted considerable attention \cite{Steppke2017,Li2022,Banerjee2018,Hameed2022,Lu2014,Hicks2025}. Examples include tuning lattice and electronic structures \cite{Baldini2015,Hicks2014}, modifying charge and magnetic orders \cite{Kim2018,Higo2022}, and inducing exotic states such as fractional topological phases \cite{Worasaran2021,Ghaemi2012}. While epitaxial strain has been used to modify lattice structure of thin film samples, uniaxial strain \cite{Landauer1955} has the benefit of explicitly breaking lattice symmetry, allowing one to continuously tune sample strain along a specific direction of materials \cite{Hicks2025}. In this regard, it turns out to be particularly useful for disentangling coexisting orders and tuning electronic band structures in quantum materials. One of the most well-known examples is cuprate high-$T_{c}$ superconductors, which display intriguing phase diagrams with multiple phases appearing along with superconductivity, such as charge and spin density waves \cite{Tranquada1996,Tranquada1997,Abbamonte2005}, pseudogap \cite{Homes1993,Timusk1999,Li2008}, and nematicity \cite{Lawler2010,DaSilvaNeto2013,Achkar2016}.

In the La$_2$CuO$_4$-based superconductors with near-1/8 hole doping, such as $\rm La_{2-x-y}Nd_{x}Sr_{y}CuO_{4}$ (LNSCO) and $\rm La_{2-x}Ba_{x}CuO_{4}$ (LBCO) \cite{Abbamonte2005,Ichikawa2000,Hucker2011,Axe1989}, it was observed that superconducting transition temperature is  strongly correlated with the charge (CSO) and spin stripe order (SSO). The superconducting transition temperature ($T_{c}$) in x=1/8 LBCO is reduced to as low as $\sim 3$~K in the presence of the static CSO, indicating the competition between the CSO and superconductivity. This was explained as arising from the peculiar structure of CSO, with stripe directions rotating 90 degrees between layers, which frustrates the interlayer Josephson coupling necessary to achieve three-dimensional superconductivity. Since uniaxial strain may directly modify the static CSO structure, it can be a particularly effective perturbation for studying the relationship between superconductivity and stripe order in these systems. Previous studies have reported the sensitivity of stripe order to uniaxial stress at near-1/8 doping, such as $\rm La_{1.88}Sr_{0.12}CuO_{4}$ (LSCO) \cite{Choi2022,Wang2022,Martinelli2024}, LNSCO (x=0.125) \cite{Boyle2021,Gupta2023} and LBCO (x=0.115, 0.125, and 0.135) \cite{Jakovac2023,Kamminga2023,Thomarat2024,Guguchia2020, Guguchia2024}.

\begin{figure*}
    \centering
    \includegraphics[width=1\linewidth]{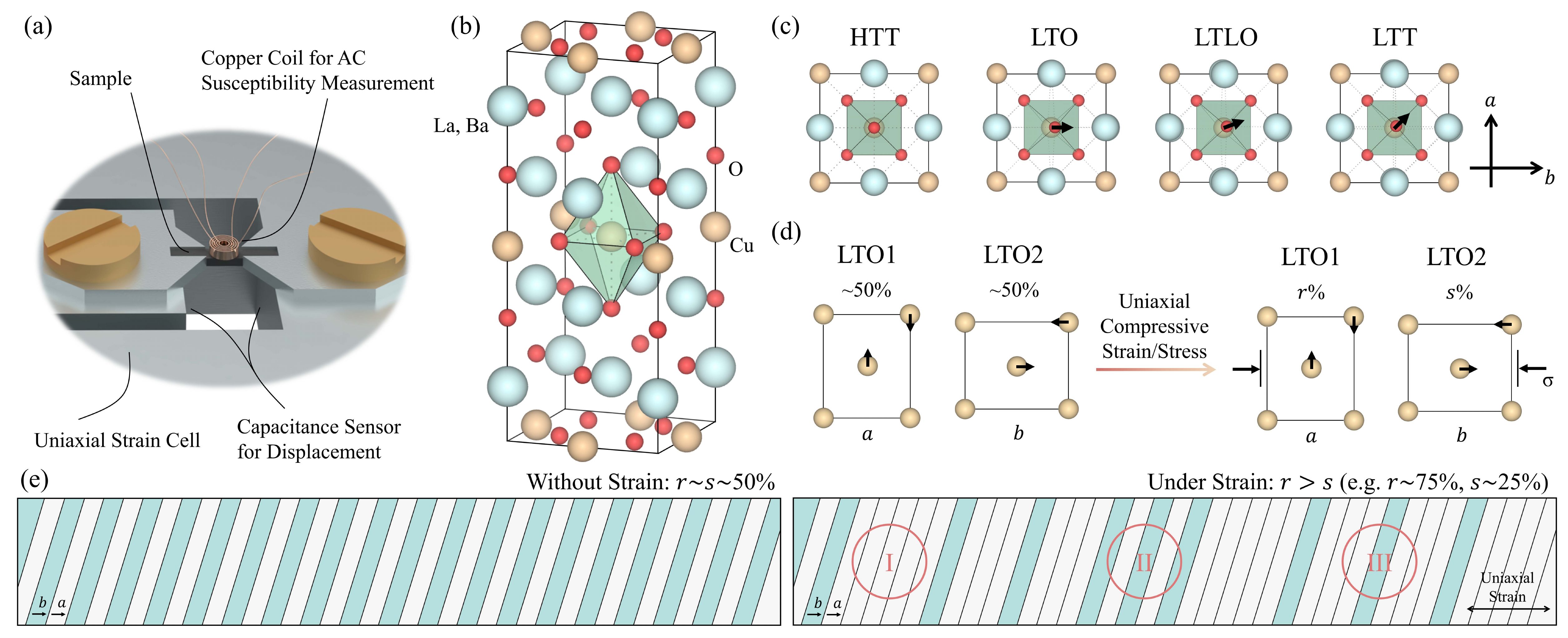}
    \caption{(a) Sketch of the strain device used to apply uniaxial compressive strain (photos of it in Fig. S1 in Supporting Information). A capacitance sensor is placed inside the gap between the two bridges for strain (displacement) measurement. (b) The three-dimensional view of the lattice structure. (c) Cu-O plane diagram showing the CuO$_6$ octahedral tilts in HTT, LTO, LTLO, and LTT phases. The black arrows denote the tilt directions. A heuristic illustration of twin domain population modulating resulting from applying uniaxial strain is shown in panels (d) and (e) for microscopic and macroscopic views, respectively. In (d), LTO1 and LTO2 denote the domain types with the lattice constants a and b along the strain direction, respectively, which are plotted in white and cyan colors in (e). The LTO domains are in layered arrangement in ab plane and elongate along c axis \cite{Lavrov2002} as shown in the left sketch in (e), where the global domain populations are about equal ($r/s \sim 1$). The right sketch shows the change of domain population under uniaxial strain, for example, $r\sim 75\%$ and $s \sim 25\%$. The I, II, and III labeling refers to the example regions with different local domain populations while uniaxial strian is applied. The region I denotes a local domain population ratio of $r_{local}/s_{local}>>1$, while the region II and III denotes a ratio of $r_{local}/s_{local}\sim1$ and $r_{local}/s_{local}>1$, respectively.}
    
\end{figure*}

Most of these studies report that the CSO and SSO tends to be suppressed under compressive uniaxial stress, while superconductivity seems to be enhanced. However, the detailed behavior of CSO, SSO, superconductivity, and structure under uniaxial stress seems to differ in these studies. For example, the CSO and SSO ordering temperature was found to be suppressed strongly in the NMR study by Jakovac et al. \cite{Jakovac2023}, while only a modest suppression of the SSO ordering temperature was observed by Guguchia et al. \cite{Guguchia2024}. This and other observed discrepancies may not be too surprising given that these studies have been carried out with various strain/stress cell technologies, which could be very sensitive to thermal expansion and the stiffness of the combined sample-epoxy-cell system. Another important consideration is the structural instability in LBCO samples with $x \sim 1/8$ doping. In these samples, the structural transition from the so-called low-temperature orthorhombic (LTO) phase to the low-temperature tetragonal (LTT) phase occurs around 55~K \cite{Axe1989,Billinge1993,Hucker2012}. This transition involves the change of the CuO$_6$ octahedral tilt direction from 45° (LTO) to 0° (LTT) with respect to the Cu-O bond direction as shown in Fig. 1. Since strain directly affects the structure, detailed understanding of structural response to uniaxial stress/strain would be important for understanding of superconductivity. 

Here we report our detailed investigation of the structural and superconducting properties of LBCO (x=1/8) sample under uniaxial strain applied along the orthorhombic a/b direction. The same cryostrain cell was used for both measurements to facilitate a direct comparison between the X-ray diffraction and AC susceptibility data. We found that moderate uniaxial strain up to 0.34\% leads the sample to separate into multiple structural phases with distinct superconducting properties. We also observed that the structural change is not reversible, indicating plastic deformation of the structure due to uniaxial strain. Our study illustrates the complexity of elucidating the role of intertwined order parameters in such a system close to structural instability.

\begin{figure*}
    \centering
    \includegraphics[width=0.85\linewidth]{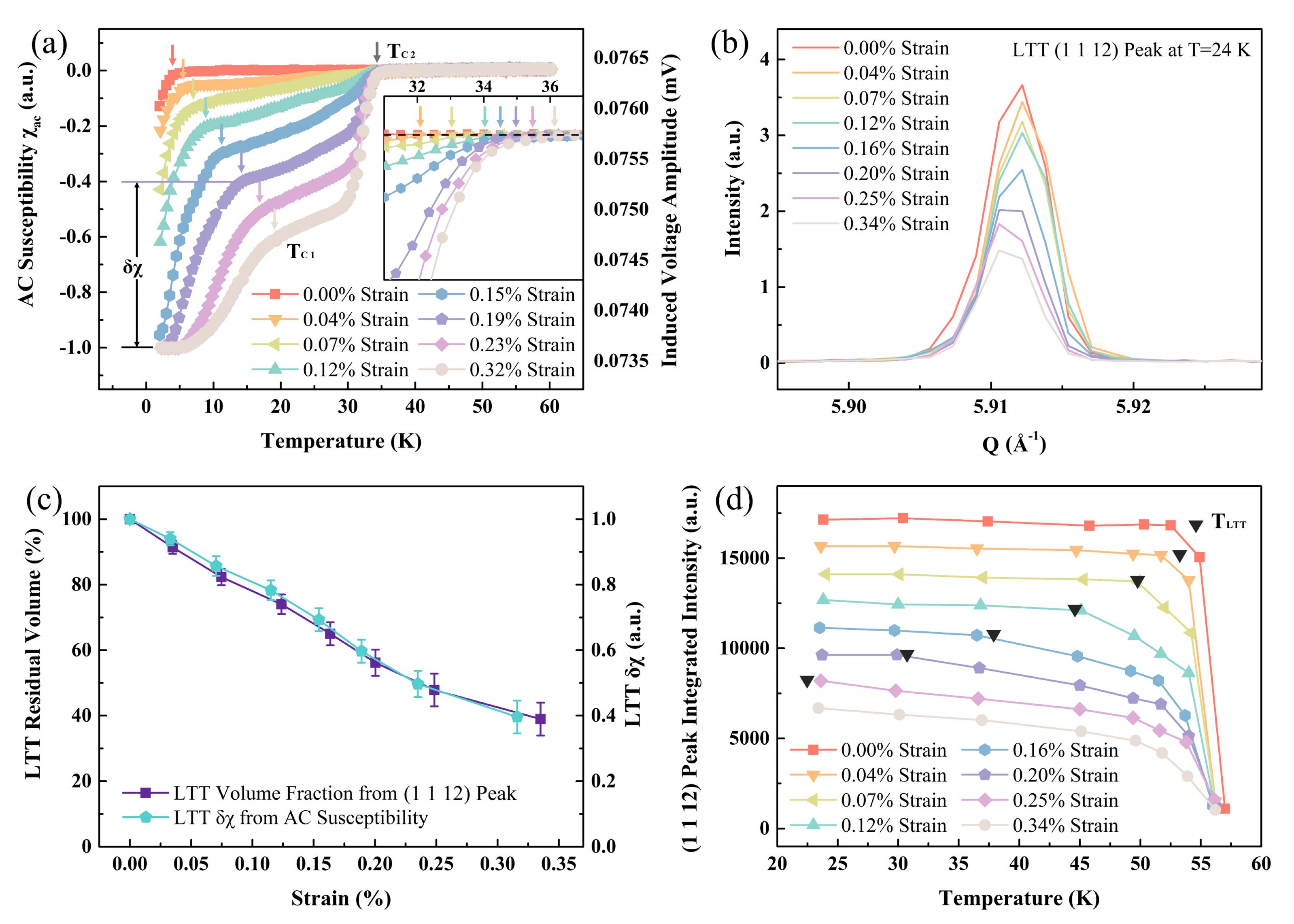}
    \caption{(a) AC susceptibility data obtained under different strains (Sample 1). The arrows indicate superconducting transition temperatures determined by the knee point method. The inset shows the magnified view around $T_{c2}$. The full scale of $\chi_{AC} \approx -1$ corresponds to the induced voltage change of $\sim 3.7\%$ following Ref.~\cite{Mark2017,manual-razorbill}. (b) The strain dependence of the LTT superlattice peak, (1 1 12), measured at 24 K from Sample 2. (c) Strain dependence of the LTT volume fraction derived from the integrated intensity of the (1 1 12) peak at 24 K (purple squares). Also plotted is the shielding fraction change $\delta \chi$ below $T_{c1}$ (cyan pentagons). (d) Temperature dependence of the integrated intensity of the (1 1 12) peak, plotted for each strain value. The black inverted triangles mark the temperatures separating two regimes with different temperature dependencies. }
  
\end{figure*}

\section{Results}

\subsection{Superconductivity Under Uniaxial Strain}

We applied uniaxial compressive strain in the LTO phase of an x=1/8 LBCO single crystal using a commercial strain device, and characterized the superconductivity of LBCO by measuring AC susceptibility as shown in Fig.~2(a) (Sample 1. For a complete list of samples with detailed thermal and strain history, see Tab. S1.). The strain is changed at 60~K for each temperature cycling to ensure that the sample is in the LTO phase when the strain change occurs. The sample is then cooled down to the base temperature, and measurements were performed. Since the main effect of the uniaxial strain is tuning the LTO twin domain population, the microscopic strain may not vary significantly. The strain is measured using the capacitive sensor of the cryostrain cell, which measures the overall change in the sample dimension. This nominal strain value will be used throughout this paper. 

Under zero strain, there is a single superconducting transition from LTT phase, $T_{c1}\sim3$ K, marked by arrows in Fig.~2(a), which is consistent with earlier results \cite{Hucker2011,Hucker2012}. With increasing strain, $T_{c1}$ increases continuously, reaching $T_{c1}\sim20$ K for $\epsilon=0.32\%$. In addition, a second superconducting transition appears at a higher temperature of about $T_{c2}\sim 32$ K as soon as a small strain of $\epsilon=0.04\%$ is applied. The magnified view near $T_{c2}$ is shown in the inset. With increasing strain, the $T_{c2}$ is slightly enhanced, reaching $\sim 36$ K for $\epsilon=0.32\%$. We note that the $T_{c2}$ value is different from both the onset of two-dimensional (2D) superconducting correlations at $\sim 40$~K and the transition to 2D superconducting order at $T_{c,2D} \sim 18$~K observed in x=1/8 LBCO samples \cite{Li2007,Tranquada2008}. The value of $T_{c2}$ is similar to $T_{c,3D}$ for maximum stress observed in earlier uniaxial stress experiments on the x=0.115 sample \cite{Guguchia2024}. It can also be seen that the superconducting transition at $T_{c1}$ broadens continuously with increasing strain, which is accompanied by a very broad superconducting transition with the onset temperature $T_{c2}$. However, for $\epsilon \geq 0.19\%$, a sharp transition at $T_{c2}$ is observed, in addition to the broad transition between $T_{c2}$ and $T_{c1}$. This observation suggests that two distinct superconducting behaviors coexist above $T_{c1}$ for $\epsilon \geq 0.19\%$. 

In addition to the changes in the transition temperatures, significant changes in the ratio of shielding fraction are observed under strain. This can be identified by the AC susceptibility changes $\delta\chi$ due to the two transitions. The shielding fraction associated with $T_{c2}$ (abrupt drop of $\chi$ at $T_{c2}$) increases with strain while the change of $\chi$ below $T_{c1}$ decreases with increasing strain, reaching a minimum value around $\delta\chi \approx 0.4$ under $\epsilon=0.32\%$.

\subsection{Lattice Structure and Phase Separation}

To understand the observed strain dependence of superconductivity, we carried out X-ray diffraction experiments to study the structure using Sample 2, but with the same strain cell and cooling history. One way to study the LTT-LTO transition is by monitoring the intensity of a superlattice peak unique to the LTT phase. In Fig. 2(b), the (1 1 12) superlattice peak intensity at the lowest temperature is plotted as a function of uniaxial strain. The peak intensity decreases with increasing strain, falling to $\sim 40\%$ of the original intensity for $\epsilon=0.34\%$. A similar peak intensity decrease under uniaxial stress was observed in x=0.115 LBCO as well \cite{Guguchia2024}. We associate this decrease in superlattice peak intensity with a reduction in the volume fraction of the LTT phase, which is expected to be disfavored by uniaxial strain that breaks tetragonal symmetry. The strain dependence of the LTT volume fraction at $\sim 24$~K is plotted in Fig.~2(c). It is interesting to note that the strain dependence of the LTT volume fraction mirrors that of the shielding fraction from $T_{c1}$ ($\delta \chi$) well, which suggests that the superconducting transition at $T_{c1}$ is associated with the LTT phase.

In Fig.~2(d), we plot the integrated intensity of the superlattice peak as a function of temperature and strain. For each strain value, one can notice three regimes exhibiting distinct temperature dependence. The peak intensity remains constant in the lowest temperature regime, and then it decreases gradually with increasing temperature. Finally, the peak intensity drops rapidly to zero, signaling the LTT to LTO transition. The grey inverted triangles denote the demarcation temperature between the constant and linear dependence, $T_{\text{LTT}}$. The appearance of this intermediate strain/temperature regime seems to suggest that moderate uniaxial strain causes structural phase separation in the sample. In their earlier studies, Guguchia and coworkers reported that low-temperature less-orthorhombic (LTLO) phase is this intermediate phase \cite{Guguchia2024}, which is confirmed in our X-ray study (see Fig. S5 in Supporting Information). Therefore, we can conclude that the LTT phase is suppressed with uniaxial strain, while the LTLO phase is enhanced, until the structure changes to the LTO phase. However, this LTT-LTLO-LTO sequence cannot account for the missing superlattice peak intensity at the lowest temperature in the LTT phase with increasing strain. This is apparent even with the smallest strain used $\epsilon=0.04\%$, and more than 50\% of the LTT intensity is missing for the maximum strain value. One possible explanation for this is the coexistence of the LTT phase with another structural `phase' under uniaxial strain. The presence of this additional structural `phase' can be surmised from the strain-temperature phase diagram as described below.

The X-ray diffraction results are summarized in the phase diagram in Fig. 3(a). For a pristine x=1/8 LBCO, there is a pure LTO to LTT transition at $T_{\text{LTO}}\sim 55$ K as reported in previous studies \cite{Hucker2011,Hucker2012}, which decreases by about 7~K with strain. The boundary $T_{\text{LTT}}$ separating region 1 and 2 (LTLO phase appearance), shows almost quadratic dependence on the uniaxial strain. We also overlay the superconducting transition temperatures determined from Fig. 2(a). We note that the $T_{\text{LTT}}$ line crosses the $T_{c2}$ line around $\epsilon \sim 0.17\%$. Since $T_{c2}$ shows a sharp superconducting transition at above $30$~K only for $\epsilon \geq 0.19\%$, the LTLO phase seems to be associated with the appearance of the sharp transition, as shown in Fig. 2(a) and discussed in previous studies \cite{Hucker2011,Guguchia2024,Hucker2012}. On the other hand, for $\epsilon < 0.17\%$, only a broad superconducting transition is observed. Since pure LTT phase ($\epsilon = 0.00\%$) has only one transition at $T_{c1}$, the broadness of the transition below $T_{c2}$, therefore, seems to be related to the new structural `phase' mentioned above. We call this a plastically deformed nano-domain structure (PDNS) for reasons discussed below.

\begin{figure*}
    \centering
    \includegraphics[width=1\linewidth]{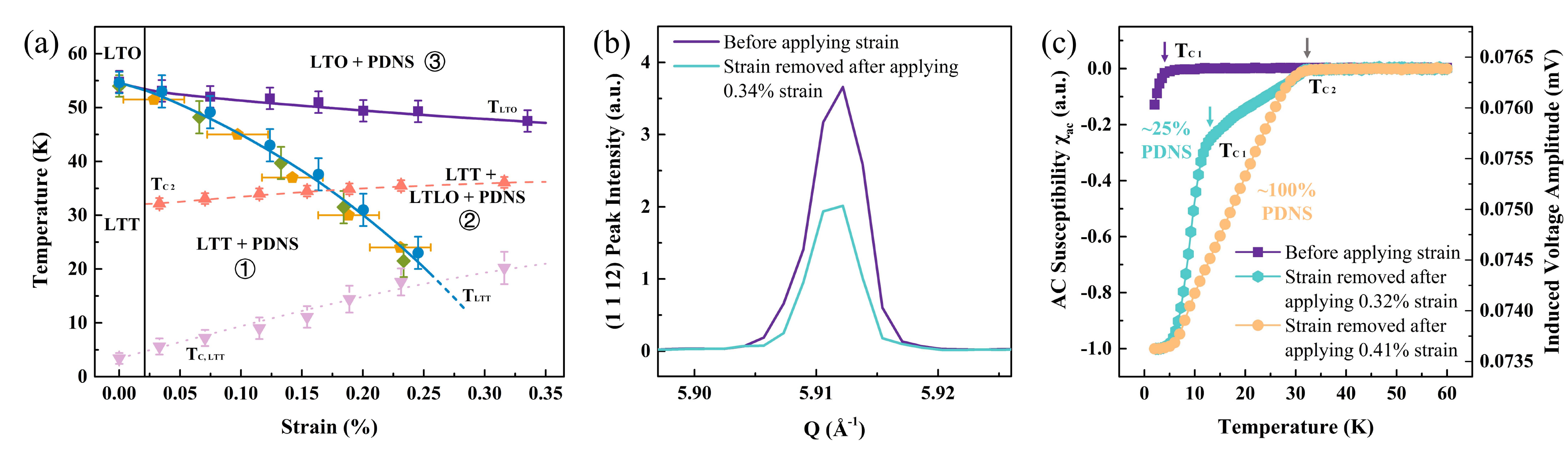}
    \caption{(a) Phase diagram of LBCO under uniaxial compressive strain. The purple squares are the LTO to LTT transition temperatures ($T_{\text{LTO}}$). The phase boundary between the LTT and LTLO structure ($T_{\text{LTT}}$) is determined from  Fig. 2(d) (blue circle), as well as from studying structural Bragg peaks (green rhombus and orange pentagon). See Fig. S4-S6 for the data. The red up triangles and pink down triangles show the superconducting transition temperatures of $T_{c2}$ and $T_{c1}$, from Fig. 2(a). The solid, dashed, and dotted lines are guides for the eyes. (b) and (c) show the (1 1 12) peak and AC susceptibility of the samples before applying strain (purple) and after removing the applied strain (cyan and orange). The strained sample in (b) (Sample 2a, Sample 2 after removing strain) is relaxed for 15 min after removing the applied voltage of strain cell at 60 K, and cooled down again for the measurement. The strained samples in (c) (Sample 1a in cyan and Sample 4a in orange, Sample 1 and Sample 4 after removing strain) are warmed up to room temperature under zero strain after removing the applied strains at 60 K. They are  fully relaxed at room temperature before cooling down again under zero strain for the measurements. See Tab. S1 for more details of sample history.}
    
\end{figure*}

\subsection{Plastic Deformation from PDNS}

All data discussed so far were obtained while increasing the strain. However, we found that the sample does not recover its original state after the strain is removed. In other words, plastic deformation of the sample results from our uniaxial strain experiments. The diffraction and susceptibility data before and after the strain studies are compared in Fig. 3(b) and 3(c), respectively. After removing the strain, the (1 1 12) LTT peak intensity does increase for Sample 2a, but is still much smaller than the intensity before the strain was applied. The broad superconducting transition $T_{c2}\sim 33$ K persists even after removing strain, below which the susceptibility decreases gradually with decreasing temperature, as seen in the intermediate temperature regime in Fig.~2(a). Such a retention of superconducting behavior after removing the applied stress is also observed in another system recently \cite{Deng2025}. The lower temperature superconducting transition $T_{c1}$ occurs at $\sim 14$~K with $\delta\chi \sim 75\%$ for Sample 1a, indicating a PDNS volume fraction of about $\sim 25\%$. We can also estimate the PDNS volume fraction from the loss of LTO volume fraction, because the LTT and LTLO phases will change back to LTO phase during the LTO-LTT transition, while PDNS will remain unchanged. The $\sim 27\%$ LTO volume fraction loss after applying 0.31\% strain (see Fig. S2) indicates that about $27\%$ of the sample volume is in the PDNS phase, consistent with the AC susceptibility estimate. 

The superconducting behavior of PDNS can be more clearly seen in the AC susceptibility data for Sample 4a, shown in Fig. 3(c). Sample 4a was subject to 0.41\% uniaxial strain, which resulted in a dominant PDNS volume fraction (see Fig. S8 and S9). Only a broad, almost linear, transition with onset temperature at $T_{c2}$ can be observed, indicating that the intermediate temperature response shown in cyan for Sample 1a can be associated with the PDNS.

To better understand the plastic deformation resulting from our cryostrain measurements, we have carried out high-resolution transmission electron microscopy (HRTEM) studies on the deformed sample. This sample is labeled Sample 1t to indicate that this is Sample 1 after strain application/removal. Experiments were done under zero strain and at room temperature. The HRTEM images of the ab plane obtained from two separate regions of Sample 1t are shown in Fig. 4(a) and 4(b), with their fast Fourier transformation (FFT) patterns in 4(c) and 4(d), respectively. In Fig. 4(a), an extended region of a square lattice in the tetragonal (HTT) phase can be seen, with a disordered structure visible in the top left corner (PDNS). Its FFT pattern in Fig. 4(c) exhibits the four-fold symmetry expected from the HTT structure. Fig. 4(b) shows a very different structure. There is no clear HTT structure, and disordered structures proliferate, which is also clear from the FFT pattern dominated by diffuse scattering intensity. However, one can find ``nano-domain"s of a few nanometers in size, marked by orange rectangles, embedded within this matrix of amorphous structure. We can also find faint peaks in the FFT pattern in Fig. 4(d), which match the lattice constants of the LTO phase.

To gain a more quantitative understanding, Fig. 4(f-g) shows electron diffraction patterns of the ab plane for Sample 1t. These should be compared with Fig. 4(e), which is the electron diffraction pattern from Sample 3, a pristine sample cut from the same crystal, which was not subject to a strain experiment. A clear four-fold symmetry of the HTT phase is observed for this sample. The lattice constant of HTT is calculated to be $a_{HTT} = 5.35$~\AA, consistent with the previous study \cite{Hucker2012}. The electron diffraction patterns obtained for two different regions of Sample 1t show clear differences, as illustrated in Fig. 4(f) and 4(g). In Fig. 4(f), the same four-fold pattern for the HTT phase can be seen, with the same lattice constants $a_{HTT}$. However, additional diffraction peaks are found in Fig. 4(g), which suggests the presence of additional domains with different orientations. The (2 0 0) and (0 2 0) peaks in Fig. 4(g) are observed to have different d spacings, corresponding to lattice constants $a_{orth} = 5.32$ \AA\ and $b_{orth} = 5.39$ \AA. These orthorhombic peaks indicate the presence of LTO nano-domains within PDNS, as illustrated in Fig. 4(b). For Sample 1t, there is pronounced diffuse intensity in both electron diffraction patterns, which is due to the scattering from disordered regions. This can be seen more clearly in azimuthally averaged intensity curves as shown in Fig. 4(h). There are three peaks marked by the grey dashed lines, which collectively resemble a glass/liquid structure factor. The first ring is located at $d_{ring} \sim 3.3$ \AA, which is close to the bond length between Cu-La (two strongest scatterers). The diffuse intensity of the FFT patterns of Fig. 4(a) and 4(b) are also plotted in 4(h), which is very similar to the electron diffraction data. Although the HRTEM data were obtained ex-situ at room temperature (above the LTO-HTT structural transition $\sim 230$~K \cite{Hucker2011}), it is unlikely that plastically deformed PDNS structure would experience the LTO-HTT structural phase transitions. The observed strong structural disorder might be the reason for the broad superconducting transition below $T_{c2}$ as shown in Fig. 3(c).

\begin{figure*}
    \centering
    \includegraphics[width=1\linewidth]{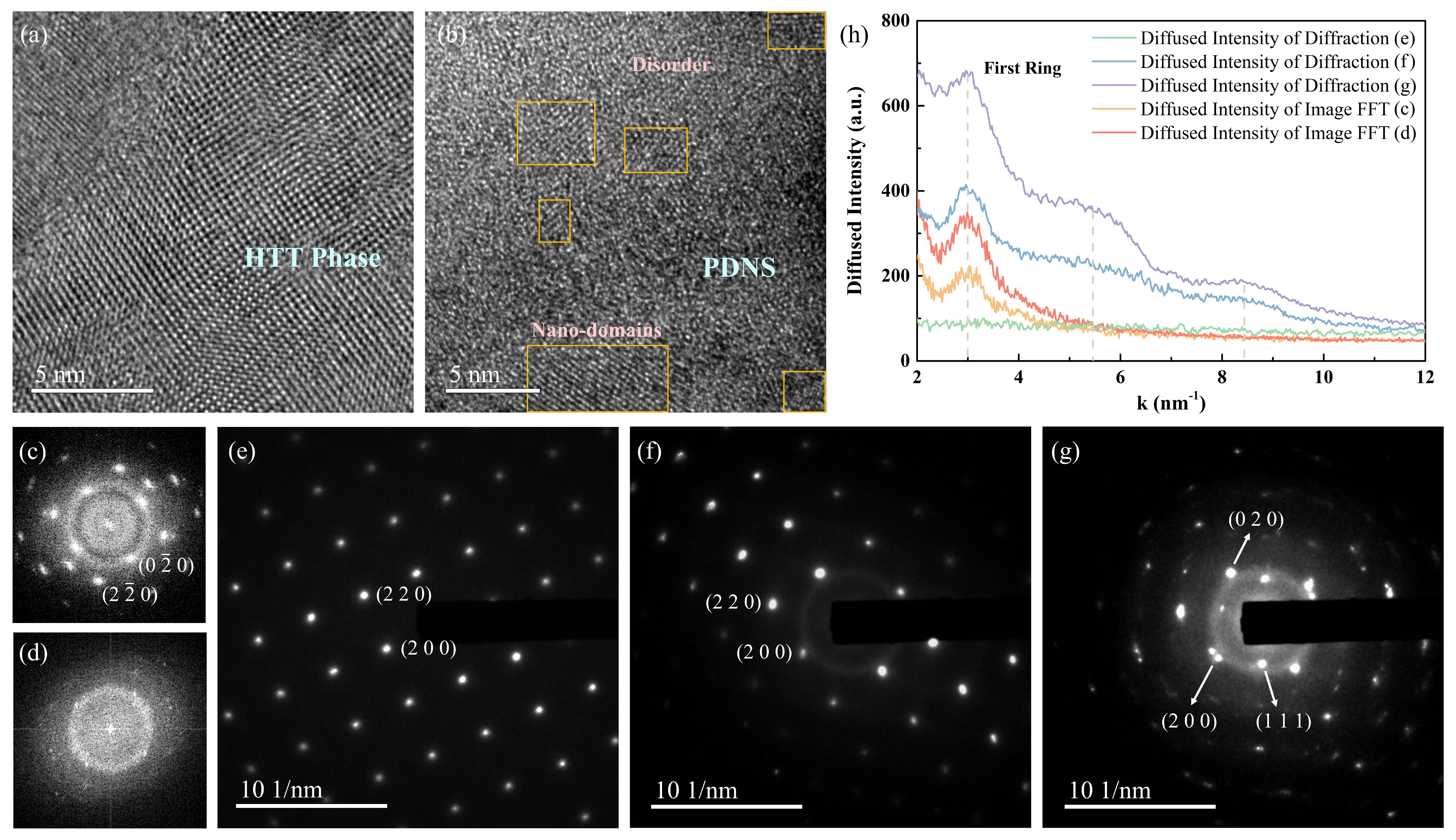}
    \caption{(a) and (b) show HRTEM images of two different regions of the strained LBCO (Sample 1t, Sample 1 taken off from strain cell after strain removal) within the ab plane. (c) and (d) show pseudo diffraction patterns from (a) and (b), respectively, obtained from Fast-Fourier Transform (FFT). (e) Electron diffraction pattern of pristine LBCO single crystal within the ab plane (Sample 3). (f) and (g) Electron diffraction patterns of Sample 1t at different locations within the ab plane. (f) is taken near the boundary of HTT and PDNS with a clear four-fold symmetry, while (g) is taken in PDNS showing an overlapping of peaks from different rotated nano-domains. Additional HRTEM data can be found in Fig. S7. (h) The azimuthally averaged intensities of the diffuse intensity of the diffraction patterns (e) to (g) and FFT patterns (c) and (d). The dashed lines mark the diffuse ring positions. Note that all the measurements are conducted at room temperature with the samples taken off from the strain device.}
    
\end{figure*}

\section{Discussion}

One of the main findings of the current study is the observation of structural phase separation below the LTO-LTT transition when the sample is cooled under uniaxial strain conditions. It is important to note that the primary response of the sample when uniaxial stress is applied at T=60 K is the change of the orthorhombic twin domain population (see Fig. S2 and S3 for strain tuning of the LTO twin domains) \cite{Choi2022,Lavrov2002,Zheng2018}. To explain the origin of the phase separation, let us consider two types of orthorhombic twin domains, labelled LTO1 and LTO2, as shown in Fig. 1(d). We define r and s as the volume fraction of LTO1 and LTO2 domain populations, respectively. The ratio r/s is expected to be around $\sim 1$ on average without strain, depicted in the left-hand side of Fig. 1(e). When uniaxial stress is applied, this ratio will change (i.e. detwinning will occur). Even for the highest strain value used in Fig. 2(a), we could not detwin the sample completely, but r/s reaches about 4 on average (see Fig. S2). However, we also expect this ratio to exhibit local fluctuations due to the inhomogeneous spatial distribution of strain in this type of uniaxial stress application, as demonstrated for a silicon wafer in Ref.~\cite{Zheng2018}. This situation is illustrated in the right-hand side of Fig. 1(e), in which we assume there are three types of local r/s ratio situations. We use $r_{local}$ and $s_{local}$ to denote the local values of the domain population to distinguish from the average values r and s. It is then reasonable to expect region II with $r_{local}/s_{local} \sim 1$ will turn into the LTT phase below the LTO-LTT transition temperature. On the other hand, region I ($r_{local}/s_{local} \gg 1$) is entirely made up of LTO1-type twin domain, which seems to “break” into small nano-domains to respond to the uniaxial stress applied to the “wrong” direction. The LTO structure persists in small pockets, with in-between regions taking on a highly glassy structure. The resulting structure is the LTO nano-domains surrounded by amorphous matrix as seen in the PDNS structure of Fig. 4(b). In between these two extremes, region II ($r_{local}/s_{local} > 1$) settles into the slightly orthorhombic LTLO structure that matches the nominal strain and lattice constant anisotropy. When the sample is fully detwinned ($r/s \gg 1$), the resultant low temperature structure will be mostly like region I, with dominant PDNS phase. Indeed, we find Sample 4 is predominantly in PDNS phase after being subject to 0.41\% strain.

The three structural phases, resulting from the phase separation caused by the uniaxial strain, exhibit distinct superconducting transitions. The PDNS phase shows a very broad transition with onset temperature near $T_{c2}$, with only a gradual (almost linear) increase of the superconducting volume fraction with cooling. This association is most clearly observed for the PDNS-dominant Sample 4, as shown in Fig. 3(c), which directly associates the broad transitions under strain with the PDNS phase. On the other hand, a sharp transition at $T_{c2}$ is only observed when the sample is highly strained ($\epsilon \gtrsim 0.17 \%$) as shown in Fig. 2(a). Since the LTLO phase only appears under the same highly strained condition as shown in Fig. 3(a), this sharp transition above 30 K is associated with the LTLO phase, as discussed in previous studies \cite{Hucker2011,Guguchia2024,Hucker2012}. Finally, $T_{c1}$ can be associated with the LTT phase, since its shielding fraction mirrors that of the LTT volume fraction as shown in Fig. 2(c).

Another interesting observation reported here is the enhanced superconducting transition temperatures with increasing uniaxial strain. In particular, $T_{c1}$, associated with the LTT phase, increases from about 3~K to about 20~K. We note that the suppression of the superconducting transition temperature in the x=1/8 LBCO sample is attributed to stripe order, which frustrates interlayer Josephson coupling \cite{Li2007}. Then, application of uniaxial strain might provide the necessary tetragonal symmetry breaking to relieve the frustration and enhance the apparent $T_{c1}$. The observed broadening of the $T_{c1}$ with increased strain indicates the presence of disorder, resulting in a series of superconducting transitions at slightly different temperatures. The fluctuation of local stress/strain discussed above could be the source of disorder, contributing to varying degrees of $T_{c1}$ suppression. This explanation is also consistent with the observed sharper $T_{c1}$ when external stress is removed as shown in Fig. 3(c), since the fluctuation of local strain would disappear in this case. 
We note that previous studies pointed out the role of disorder in the suppression of charge density waves and enhancement of superconductivity \cite{Campi2015,Straquadine2019,Xiao2021}.
However, direct investigation of CSO or SSO is required to draw any firm conclusion regarding the role of stripe order in enhancement and suppression of superconductivity. 

To understand our findings in the context of other uniaxial stress/strain experiments \cite{Choi2022,Wang2022,Martinelli2024,Boyle2021,Gupta2023,Jakovac2023,Kamminga2023,Thomarat2024,Guguchia2020,Guguchia2024}, we would like to point out that our `uniaxial strain' experiments were carried out by changing the sample dimension, which may be different from applying a constant uniaxial pressure as done in some other studies\cite{Choi2022,Gupta2023,Jakovac2023,Guguchia2024}. In the latter experiments, sample strain could change due to the change of Young's modulus and stiffness during temperature change or structural phase transitions. However, the fluctuation of local stress on structural domains during LTO-LTT transition could still be present in this case, arising from the possible variation of local LTO domain population. We note that a similar Bragg peak intensity decrease was also observed with uniaxial stress in another diffraction study of x=0.115 LBCO \cite{Guguchia2024}.

Although we associate the decrease of the LTT superlattice peak intensity with reduced volume fraction of the LTT structure, we consider a possible alternative explanation here. The superlattice peak intensity could represent the order parameter of the LTT structure, the octahedral tilting along the (110) crystallographic direction (See Fig. 1(c)). Therefore, suppression of the LTT peak could be due to a continuous decrease of the octahedral tilt angle. This scenario would suggest a second-order nature of the LTT-LTO transition. However, this transition is known to be a first-order transition accompanied by an abrupt change of the octahedral tilt angle, at least at zero stress/strain. In addition, our AC susceptibility and HRTEM data provide strong support for the volume fraction origin of the observed LTT superlattice peak intensity suppression.

\section{Conclusion}
We report our X-ray diffraction, electron microscopy, and AC susceptibility investigations of crystal structure and superconductivity when a $\rm La_{1.875}Ba_{0.125}CuO_{4}$ crystal is subject to compressive uniaxial strain along the orthorhombic a/b direction. In our experiment, uniaxial pressure was applied in the LTO phase at 60~K, forcing rearrangement of the orthorhombic twin domain population. When the sample is cooled through the structural transition temperature while constraining the sample dimension (nominally uniaxial strain), the crystal structure exhibits phase separation into three distinct structures with corresponding superconducting behavior. The superconducting transition temperature of the LTT structure is enhanced from 3~K to 20~K, but broadens significantly, with increasing uniaxial strain. This observation is accompanied by the appearance of a sharp transition around 32-36~K, which is associated with the LTLO structure. In addition, we observe that part of the sample becomes permanently disordered and exhibits very broad superconducting transition. We call this third phase a plastically-deformed nano-domain structure (PDNS) comprised of few-nanometer-sized orthorhombic nano-domains within amorphous matrix, which is revealed in our High-Resolution TEM studies. Superconductivity survives in PDNS, even PDNS dominates in LBCO under 0.41\% uniaxial strain. Our study illustrates the structural complexity associated with studying samples near the structural instability using uniaxial stress or strain.

\section{Experimental Section}

The $\rm La_{1.875}Ba_{0.125}CuO_{4}$ single crystal used in this paper is synthesized by the traveling solvent floating-zone method and studied in Ref.~\cite{Kim2008}. It is cut into suitable size along orthorhombic a axis using a wire saw, and mounted on a uniaxial strain device for the measurements. A typical sample is in a cuboid shape with widths of 0.15-0.3~mm and lengths of 1-2~mm. The applied uniaxial strain creates the PDNS during all our experiments. Therefore, several LBCO samples are cut from the same crystal, and each is used for one strain experiment in this paper.

The uniaxial compressive strain is applied by a commercial strain cell (model CS100 from Razorbill \cite{Steppke2017,Hicks2014,Hicks12014}), with its sketch shown in Fig. 1(a) (photos in Fig. S1). Under applied DC voltages to the piezoelectric stacks, the stacks drive the two sample bridges moving relative to each other and an in-situ stress is therefore applied to the sample. The sample strain can be measured by a built-in capacitance sensor between the two bridges.

Strain is applied at 60 K in the LTO phase by changing the voltages applied to the piezoelectric stacks of the strain cell, and the voltages were kept during temperature cycling. After each voltage change, the sample is allowed to relax for 15 min, followed by cooling down to the base temperature of 2 K (AC susceptibility) and about 20 K (X-ray). Both measurements are carried out during the warming cycle. The next strain step occurs when the sample reaches 60 K. The key point for this procedure is to control the population of the twinning domains in the LTO phase by uniaxial stress \cite{Lavrov2002,Gugenberger1994,Lavrov2001}. The twin domains are in layered arrangement in ab plane and propagate along the c axis as shown in Fig. 1(e) \cite{Lavrov2002}. Along strain direction, the ratio of unit cells in LTO1 and LTO2 domains is expected to be close to the ratio of domain population. And thus, the ratio change of domain population driven by uniaxial strain will result in a compression of sample length due to the difference between LTO lattice constants $a_{LTO}$ and $b_{LTO}$. Since domain population tuning is sensitive and does not change the lattice constant, a small stress may achieve a relatively large sample dimension change in our case.

The AC susceptibility is measured in Physical Property Measurement System down to 2 K using a commercial Cu coil (model MTD100 from Razorbill \cite{manual-razorbill}), and the corresponding superconducting transition temperatures ($T_{c1}$ and $T_{c2}$) are determined by knee point method. A similar technique has been tested and used in the previous studies for superconducting transitions of $\rm Sr_{2}RuO_{4}$ \cite{Steppke2017,Hicks2014}. The coil is attached on top of the sample surface by a very thin layer of GE Vanish. A current of 5 mA amplitude and 1000 Hz frequency is used to apply an alternating magnetic field to the sample, and the ac susceptibility signal is measured from the induced voltage amplitudes from the sample. A $\sim 3.7\%$ change of the induced voltage amplitudes is observed from our Sample 1, which is in a cuboid shape with 0.15 mm $\times$ 0.15 mm $\times$ 1~mm. The full range of the superconducting transition from 0.0764 mV to 0.0736 mV is also noted in Fig. 2(a). This signal matches the finite element method simulation provided by the manufacturer very well \cite{Mark2017,manual-razorbill}, which gives credence to the superconducting fraction discussed in this study.

The X-ray diffraction measurement is carried out at the BXDS-IVU beamline of Canadian Light Source. The photon energy of 19.2 keV is selected by a double crystal Si (1 1 1) monochromator. A Lambda area detector is used without analyzer crystals. The peak intensity is integrated over the whole Bragg peak of interest. The strain cell is mounted on the cold finger of closed-cycle cryostat (ARS DE202). The lowest temperature of this setup could reach about 20 K.

The electron diffraction and imaging study is conducted by Hitachi high-resolution transmission electron microscopy (HRTEM) (model HF-3300) at 300 keV at room temperature. The samples are cut by Hitachi focused ion beam (FIB) (model NB5000) down to thickness of about 100 nm. The electron diffraction patterns are taken at selected positions of about 80 nm in diameter. Three samples are measured under zero stress, including two strained LBCO after AC susceptibility measurements (used in Fig. 2(a) and 3(c)) and one pristine LBCO for comparison.

\medskip
\textbf{Supporting Information} \par 
Supporting Information is available from the Wiley Online Library or from the author.

\medskip
\textbf{Acknowledgments} \par
Work at the University of Toronto was supported by the Natural Sciences and Engineering Research Council (NSERC) of Canada through the Discovery Grant No. RGPIN-2025-06514, Canada Foundation for Innovation, and Ontario Research Fund. Research performed at the Canadian Light Source is supported by the Canada Foundation for Innovation, NSERC, the University of Saskatchewan, the Government of Saskatchewan, Western Economic Diversification Canada, the National Research Council Canada, and the Canadian Institute of Health Research.
B.G. acknowledges supports from the CLS Graduate and Post-Doctoral Student Travel Support Program and from the US National Science Foundation (NSF) Grant Number 2201516 under the Accelnet program of Office of International Science and Engineering (OISE). Work at Brookhaven is supported by the Office of Basic Energy Sciences, Materials Sciences and Engineering Division, U.S.\ Department of Energy (DOE) under Contract No.\ DE-SC0012704. We also acknowledge Open Center for Characterization of Advanced Materials (OCCAM) in University of Toronto for providing access to the TEM study.

\medskip
\textbf{Conflict of Interest} \par 
The authors declare no conﬂict of interest.

\medskip
\textbf{Author Contributions} \par 
Y.-J. K. and B. G. conceived and designed the experiments. G. D. G. and J. M. T. synthesized the samples. B. G. prepared the samples for all the strain experiments. B. G., G. J., C. H. and Z. S. conducted X-ray studies with the help from N. A., B. D. M. and A. R.. B. G. conducted AC susceptibility measurements. E. N. and B. G. prepared the HRTEM samples and conducted HRTEM characterizations. B. G. and Y.-J. K. analysed the data, plotted the figures, and wrote the paper.

\medskip
\textbf{Data Availability Statement} \par 
The data that support the ﬁndings of this study are available from the corresponding author upon reasonable request.

\medskip
\textbf{Keywords} \par 
Uniaxial Strain, Phase Separation, Unconventional Superconductor.

\nocite{*}

\bibliography{main}

\end{document}